# Improving Prognostic Performance in Resectable Pancreatic Ductal Adenocarcinoma using Radiomics and Deep Learning Features Fusion in CT Images

**Authors:**

| # | Name | Affiliations |
|---|------|--------------|
| 1 | Yucheng Zhang | 1,2 |
| 2 | Edrise M. Lobo-Mueller | 3 |
| 3 | Paul Karanicolas | 4 |
| 4 | Steven Gallinger | 2 |
| 5 | Masoom A. Haider | 1,2,5 |
| 6 | Farzad Khalvati | 1,2 |

**Affiliations**

1: Department of Medical Imaging, University of Toronto, Toronto, ON, Canada

2: Lunenfeld-Tanenbaum Research Institute, Sinai Health System, Toronto, ON, Canada

3: Department of Radiology, Faculty of Health Sciences, McMaster University and Hamilton Health Sciences, Juravinski Hospital and Cancer Centre, Hamilton, Ontario, Canada

4: Department of Surgery, Sunnybrook Health Sciences Centre, Toronto, ON, Canada.

5: Sunnybrook Research Institute, Toronto, ON, Canada

**Corresponding Author:** Dr. Farzad Khalvati, M.A.Sc., Ph.D.

Staff Scientist, Lunenfeld-Tanenbaum Research Institute (LTRI), Sinai Health System

Assistant Professor, Medical Imaging, University of Toronto

Farzad.Khalvati@utoronto.ca

**Abstract**

As an analytic pipeline for quantitative imaging feature extraction and analysis, radiomics has grown rapidly in the past a few years. Recent studies in radiomics aim to investigate the relationship between tumors imaging features and clinical outcomes. Open source radiomics feature banks enable the extraction and analysis of thousands of predefined features. On the other hand, recent advances in deep learning have shown significant potential in the quantitative medical imaging field, raising the research question of whether predefined radiomics features have predictive information in addition to deep learning features. In this study, we propose a feature fusion method and investigate whether a combined feature bank of deep learning and predefined radiomics features can improve the prognostics performance. CT images from resectable Pancreatic Adenocarcinoma (PDAC) patients were used to compare the prognosis performance of common feature reduction and fusion methods and the proposed risk-score based feature fusion method for overall survival. It was shown that the proposed feature fusion method significantly improves the prognosis performance for overall survival in resectable PDAC cohorts, elevating the area under ROC curve by 51% compared to predefined radiomics features alone, by 16% compared to deep learning features alone, and by 32% compared to existing feature reduction methods for a combination of deep learning and predefined radiomics features.

**Introduction**

Radiomics features are designed to decode the predictive information in medical images for cancer patients. As a quantitative approach, radiomics involves the extraction and analysis of quantitative medical imaging features and establishing correlations between these features and clinical outcomes such as patient survival[1–5]. Several radiomic features have been found to be significantly associated with various clinical outcomes in multiple cancer sites such as lung, pancreas, and kidney[2,6–12].

In the past few years, the pipeline for traditional radiomics analysis has been established[1,2,9,13]. The traditional pipeline consists of four steps: image acquisition, region of interest (ROI) segmentation, feature extraction, and predictive model building. The core of the traditional radiomics studies relies on the extraction of a set of engineered features based on predefined mathematical formulas[14]. These engineered features, which are extracted from ROIs annotated by clinicians, have been designed to capture different characteristics of images[15]. For example, the first order features measure the distribution of pixel intensities while second-order features based on grey-level co-occurrence matrix (GLCM) extract texture information[14]. Efforts have been made to standardize the feature banks by implementing open source libraries such as PyRadiomics[15]. In these feature banks, thousands of engineered features from different classes of features can be extracted from 2D or 3D medical images[15]. These features can then be tested for associations with clinical outcomes such as overall survival, recurrence, or genetic mutations[8,16–18]. Several cross-cohort and multi-centre studies have shown that serval PyRadiomics features are robust to different scanners and clinician annotations[8,15,19,20].

Despite recent progress, traditional radiomics analytics pipeline has few drawbacks. First, the formulas of features are predefined, and thus, can be very similar to one another. This leads to high correlations among different radiomic features. As a result, if a feature was found to be significantly associated with a certain clinical outcome, other features with high correlation with the feature will more likely be significant as well. Consequently, while the high dimension of features increases the complexity and computational power requirements, there is no corresponding increase in the prognosis performance. Second, testing radiomic features one by one increases the chance of producing at least one false positive. Previous publications have pointed out that several radiomics studies lacked multiple testing control and hence, some discovered significant features may be the result of type I errors[21,22]. These shortcomings in the traditional radiomic analytics pipeline have inspired new research which takes advantage of the recent impressive progress in deep learning and convolutional neural networks to improve the performance of the predictive models.

Convolutional neural networks (CNNs) are one of the most frequently used deep learning architectures in computer vision tasks[23]. CNNs apply a series of convolution operations on input images, preserving the spatial relationship between pixels and mapping these relationships on to outputs. During the training phase, parameters of the convolution operations are tuned. Consequently, convolution layers can capture information specifically related to the classification task (e.g., outcome prediction) at hand. In medical imaging, this allows to generate customized feature maps for specific modality or diseases, which further improves performance[24,25]. However, training CNN parameters requires a large sample size, which is usually not available in a typical medical imaging research setting. To overcome this limitation, transfer learning-based feature extraction has been proposed[26–28].

Transfer learning was developed based on an assumption that the structures of CNNs are similar to the mechanism of human visual cortex[23,29]. The top layers of CNNs can extract general features from images, while the deeper layers are more specific to the target[23]. Pretraining CNNs using large image datasets such as ImageNet helps the model to learn how to extract general features[30,31]. Since many image recognition tasks are similar, the top layers of the network can be transferred to another target domain[27]. On the other hand, deeper layers of CNNs can extract "higher-order" information which is associated with the target outcome. Thus, if the target domain is similar to the pretrained domain, deeper layers can also be transferred to extractor features[26,32].

Deep learning and transfer learning-based feature extraction have shown promising results in cancer assessment[32–34]. Furthermore, it has also been shown that, combining predefined features with deep learning-based features further improved the performance in the prognosis of Glioblastoma Multiforme[32]. Hence, it is crucial to develop a feature reduction method which can fuse the predictive power of deep radiomics with predefined radiomic features to achieve optimal performance.

Traditional feature reduction methods can be classified into two groups: supervised (e.g., Boruta, Cox Proportional Hazard Model) and unsupervised feature reduction (e.g., PCA, ICA)[35–37]. The main difference between the two is that unsupervised methods reduce features based on the characteristics of features regardless of the outcome. In contrast, supervised methods rely on the association between features and the outcome[9,38,39]. Although previous studies compared the performance of these feature selection and reduction methods, these comparisons were made only for radiomics feature banks[9,38,39]. To further improve the performance through deep transfer learning, there is an urgent need to investigate optimal feature selection and reduction methods for a combination of deep learning and predefined radiomics feature banks in the quantitative medical imaging field.

In this paper, first, we compared the performance of three feature reduction methods: PCA, Boruta, and Cox Proportional Hazards Model (CPH). These feature reduction methods were applied to the combined feature set of predefined radiomics and transfer learning-based deep radiomic features. Next, we proposed a feature fusion method using risk scores and compared its performance to the abovementioned feature reduction methods. Our results illustrated that the proposed risk-score based feature fusion method significantly improved the performance of the prognostication of overall survival of resectable PDAC patients when compared to traditional feature reduction models (PCA, Boruta, and CPH).

**Methods**

**Dataset**

Two cohorts from two independent hospitals consisting of 30 and 68 patients were enrolled in this retrospective study. All patients underwent curative intent surgical resection for PDAC from 2007 – 2012 and 2008 – 2013 for cohort 1 and cohort 2, respectively, and they did not receive other neo-adjuvant treatment. Contrast-enhanced CT images were obtained pre-operatively. Overall survival data were collected as the primary outcome. To exclude the effect of post-operative complications on the prognosis, the patients who died within 90 days after surgery were excluded. Institutional review board approval was obtained for this study from both institutions. An in-house developed region of interest (ROI) contouring tool (ProCanVAS) was used by an experienced radiologist[40]. The reader contoured the ROIs blind to the outcome. A cohort with 68 patients from one institution was used as the training cohort while another cohort with 30 patients from a different institution was used as the test cohort.

**Radiomics Feature Extraction**

Pre-defined radiomic features were extracted using the PyRadiomics library (version 2.0.0) in Python[15]. To ensure that features were extracted from tumor regions exclusively, voxels with

Hounsfield unit (HU) < -10 and > 500 were excluded to eliminate fat and stents from the feature values. In total, 1,428 radiomic features were extracted for both cohorts. Details of the extracted features are listed in Table 1.

Table 1: Number of features extracted for different feature classes and image filters

| Filter/ Features | firstorder | glcm | gldm | glrlm | glszm | ngtdm | shape |
|---|---|---|---|---|---|---|---|
| exponential | 16 | 0 | 11 | 12 | 7 | 0 | 0 |
| gradient | 18 | 23 | 14 | 16 | 16 | 5 | 0 |
| lbp | 56 | 0 | 44 | 48 | 28 | 0 | 0 |
| logarithm | 18 | 23 | 14 | 16 | 16 | 5 | 0 |
| original | 18 | 23 | 14 | 16 | 16 | 5 | 11 |
| square | 18 | 23 | 14 | 16 | 16 | 4 | 0 |
| squareroot | 18 | 23 | 14 | 16 | 16 | 5 | 0 |
| wavelet | 144 | 184 | 112 | 128 | 128 | 39 | 0 |

**Transfer Learning Feature Extraction**

We used two transfer learning models using Lung CT pretrained and ImageNet pretrained ResNet (LungRes and ImgRes)[41]. Through adding direct paths, the ResNet model avoids the gradient vanishing problem and achieves better performance. ResNet50 in Keras library was chosen since it is a state-of-art deep learning architecture with high classification performance[42].

Two datasets were used to pretrain the ResNet model. The first dataset is ImageNet[43,44], which contains 14,197,122 natural images from 21,841 different categories. The second dataset is Non-Small Cell Lung Cancer (NSCLC) dataset, which was published on Kaggle with CT images from 888 patients[45]. ImageNet pretrained ResNet was directly available in Keras 2.0 which is a python-based deep learning library[42,46]. The LungRes CNN was trained from scratch using the lung CT images.

The process of transfer learning varies depending on the similarity of the pretrained domain and target domain. Since our target domain (pancreatic CT) is small and different from the first pretrained domain (ImageNet - natural images), during transfer learning process using ImgRes, features were extracted from a shallower layer (12$^{th}$ layer). For LungRes, since the pretrained and target domains are rather similar (lung and pancreatic CT), features were extracted from the final layer before the classifier. In total, 2048 ImgRes features and 64 LungRes features were extracted.

**Correlation**

To investigate the correlation between the features extracted using traditional radiomics pipeline (PyRadiomics) and transfer learning approaches (ImgRes, and LungRes), Pearson correlation coefficients were calculated for each pair of feature sets in the training cohort (n=68). Mean correlation coefficient was calculated for each combination of the three different feature extraction methods (PyRadiomics, ImgRes, and LungRes). The distributions of the correlation coefficients were also calculated.

**Proposed Prognosis Model**

To investigate the optimal feature reduction and fusion methods, we trained four prognosis models using CT images from Cohort 1 (n=68) and validated them in Cohort 2 (n=30). Figures 1-A, 1-B, and 1-C shows the prognosis models using three traditional feature reduction algorithms; PCA, CPH, and Bortua. In each model, the three feature banks (PyRadiomics,

ImgRes, and LungRes) were concatenated together. Then, the feature reduction algorithm was applied to these features. The remaining features were used to train a Random Forest classifier using the training cohort, with the derived model validated in the test cohort which was collected in an independent hospital site. For the CPH method, p-value < 0.05 was used as a feature selector.

Our proposed risk score-based method is illustrated in Figure 1-D. First, using the training cohort, three different Random Forest classification models were trained separately using each of the three feature banks (PyRadiomics, ImgRes, and LungRes)[47]. Each of these models was then used to produce the probability for every patient in the training cohort through 10-fold cross-validation. In the test cohort, for each patient, three probabilities were generated using the three models. We treated these probabilities as new features, based on which, the final prognosis model was built through another Random Forest classifier.

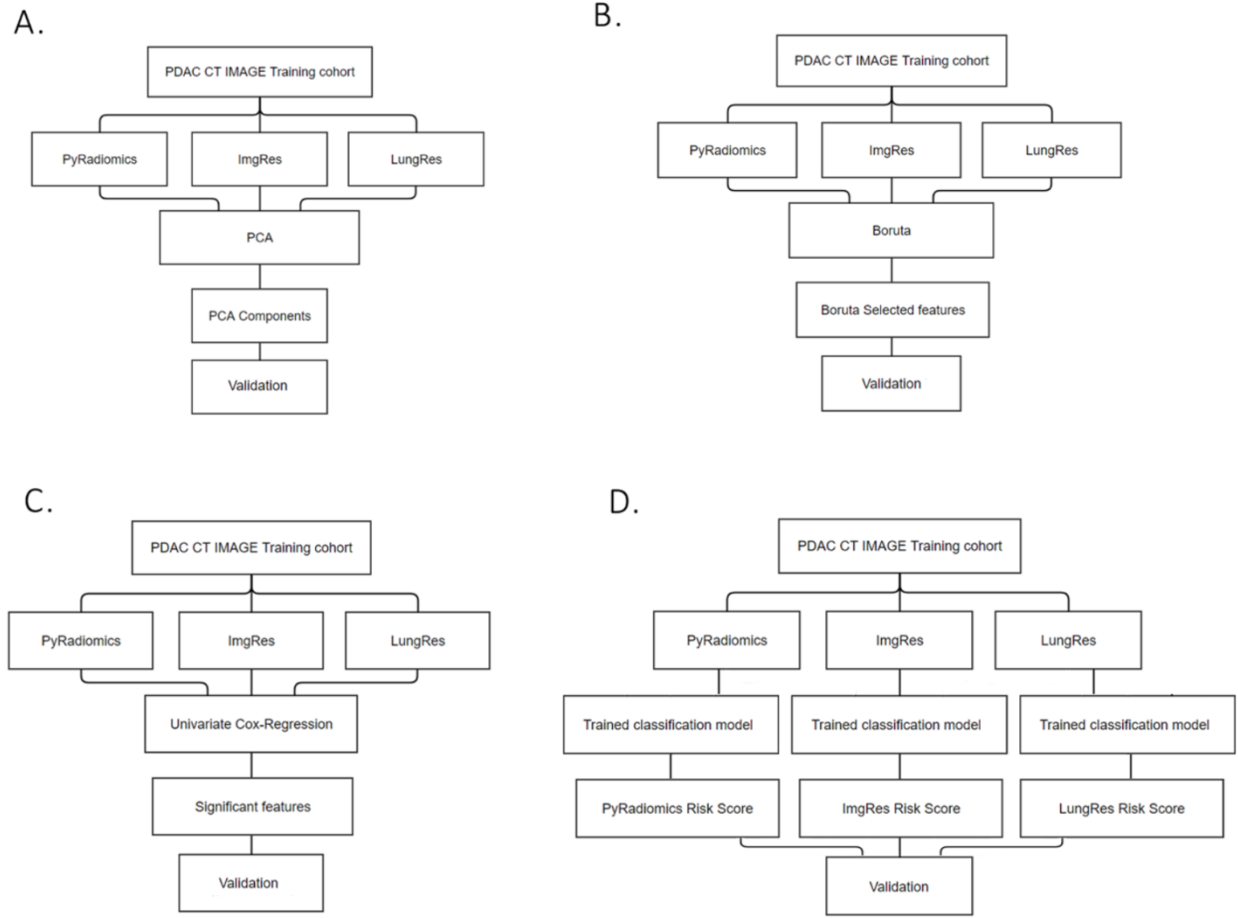

**Figure 1**. Pipelines for different feature fusion methods.

A. Unsupervised feature fusion using PCA. Features from three feature banks were fused using PCA, generating few components. Next, these components were used to build a model (Random Forest) in the training cohort. In the end, the performance of the model was evaluated in the validation cohort.

B. Supervised feature reduction using Boruta. Boruta identified prognostic features which were then used to build a prognosis model (Random Forest) in the training dataset. The model's performance was validated in the validation cohort.

C. Supervised feature reduction using Cox-Regression. Each feature was tested using univariate Cox-regression. Significant features were then used in building a prognosis model (Random Forest), which was validated in the validation cohort.

D. The proposed risk-score based feature fusion method. Three prognosis models (Random Forest) were built using features from three feature banks. The prediction outputs of these models were considered as risk-scores. Hence, for every patient, there were three risk-scores. Next, another model (Random Forest) was trained using these risk-scores in the training set and validated in the validation cohort.

Area under ROC curve (AUC) was used to measure the performance of these four approaches[48]. The ROC-based specificity tests were applied to test the difference between the AUCs of different models (DeLong et al., 1988). These analyses were performed through "pROC" package in R (Version 3.5.1)[50].

## Results

**Correlation Analysis Between Predefined and Deep Radiomic Features**

Within each feature bank, the average absolute value of Pearson correlations coefficients of 1,428 PyRadiomics features was 0.27, while ImgRes (2,048 features) and LungRes (64 features) had mean correlations of 0.23 and 0.24, respectively. This showed that PyRadiomics features had a higher correlation among each other compared to deep radiomic features. The cross-correlation of PyRadiomics and ImgRes features yielded a mean absolute coefficient of 0.15, which was similar for PyRadiomics vs. LungRes features (coefficient = 0.16). The two deep transfer learning-based features banks (ImgRes and LungRes) had a slightly higher mean correlation coefficient of 0.18. Table 2 summarizes the correlation results.

**Table 2**: Absolute Pearson correlation coefficient between features from each feature extraction method

|  | PyRadiomics (1428) | ImgRes (2048) | LungRes (64) |
|---|---|---|---|
| PyRadiomics (1428) | 0.27 | 0.15 | 0.16 |
| ImgRes (2048) | 0.15 | 0.23 | 0.18 |
| LungRes (64) | 0.16 | 0.18 | 0.24 |

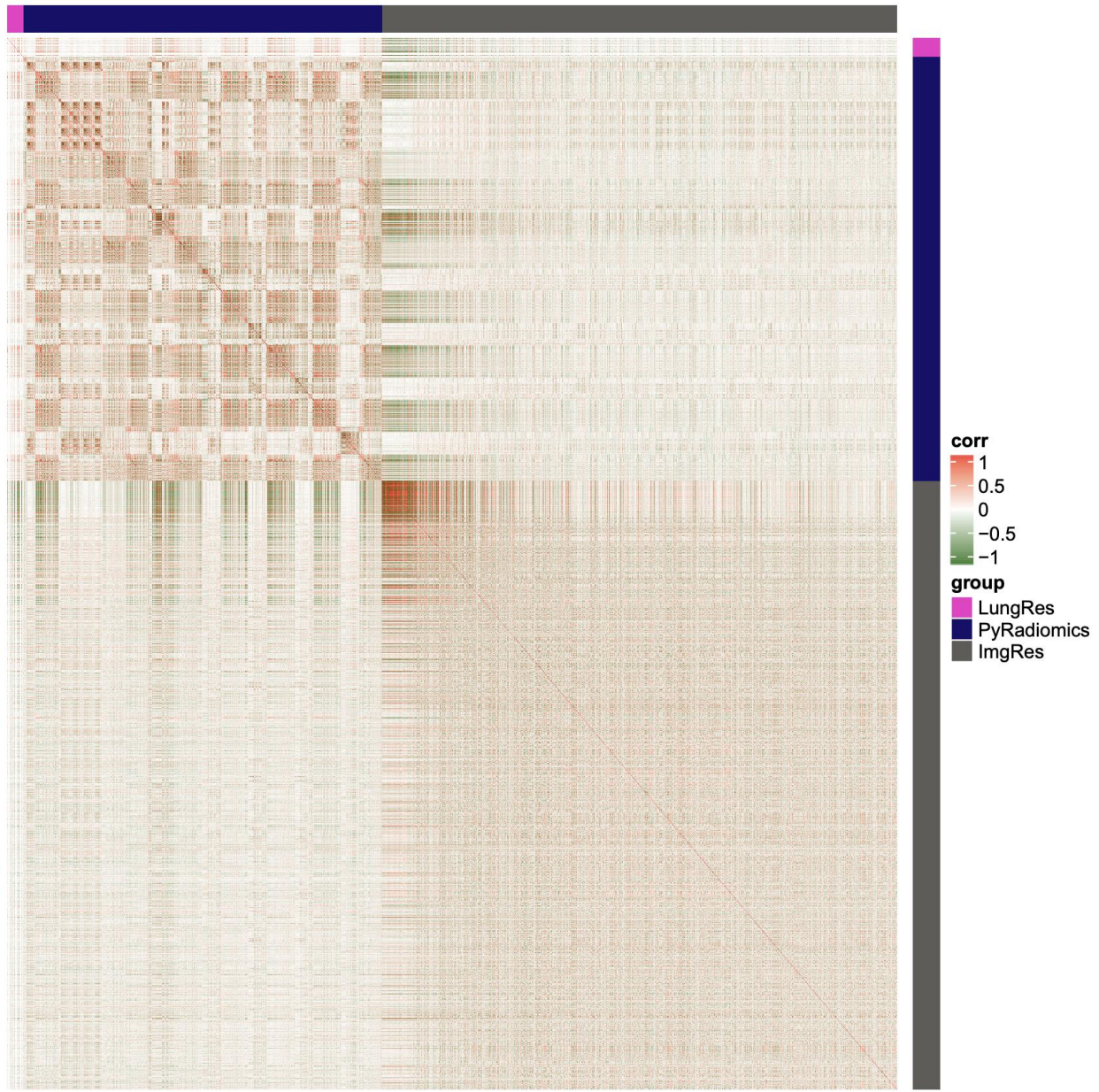

**Figure 2**. Correlation heatmap of three different feature extraction methods.

The heatmap in Figure 2 shows the correlation details. Each dot in Figure 2 represents a correlation coefficient. White color indicates that the coefficient is 0, while red and green dots represent positive or negative correlations, respectively. There were several bright color blocks in PyRadiomics versus PyRadiomics region, indicating high correlations among the

PyRadiomics features. Although ImgRes features had lower average absolute correlation coefficient, there was still a bright color block, suggesting that some ImgRes features had high correlations with each other. Despite few darker bands, the color was lighter in ImgRes vs. PyRadiomics, and LungRes vs. PyRadiomics regions, indicating that the average correlation coefficients were lower across these feature banks.

The distribution of the correlation coefficients (in absolute value) are also displayed in histogram form in Figure 3 for PyRadiomics vs. ImgRes and PyRadiomics vs. LungRes. As illustrated by skewed distributions, most of the predefined and deep radiomic features had no or weak correlation among each other. However, it was clear that some features have high correlations with coefficients higher than $0.7$[51]. This result indicates that, some deep transfer learning features (deep radiomic features) could resemble properties of certain predefined radiomic features. As an example, LungRes feature "v1" had a correlation coefficient of 0.93 with PyRadiomics feature "gradient_glcm_ClusterProminence", and 0.91 with "gradient_glcm_ClusterShade". According to PyRadiomics, these two features measure the skewness and asymmetry, and uniformity of the GLCM[15]. More details for the correlation between PyRadiomics and Transfer Learning features can be found in Table 3 and 4, where the average absolute values of correlation coefficients were calculated for each type of filter and feature.

**Table 3**. Mean absolute correlation coefficients between PyRadiomics and ImgRes features across different types of filters and features.

|  | firstorder | glcm | gldm | glrlm | glszm | ngtdm | shape |
|---|---|---|---|---|---|---|---|
| exponential | 0.15 |  | 0.16 | 0.16 | 0.14 |  |  |
| gradient | 0.19 | 0.18 | 0.16 | 0.15 | 0.19 | 0.20 |  |
| lbp | 0.12 |  | 0.16 | 0.16 | 0.15 |  |  |
| logarithm | 0.13 | 0.12 | 0.13 | 0.13 | 0.13 | 0.12 |  |
| original | 0.17 | 0.16 | 0.15 | 0.15 | 0.16 | 0.20 | 0.18 |
| square | 0.18 | 0.24 | 0.20 | 0.21 | 0.21 | 0.22 |  |
| squareroot | 0.15 | 0.14 | 0.14 | 0.15 | 0.14 | 0.16 |  |
| wavelet | 0.17 | 0.14 | 0.15 | 0.15 | 0.14 | 0.16 |  |

**Table 4**. Mean absolute correlation coefficients between PyRadiomics and LungRes features across different types of filters and features.

|  | firstorder | glcm | gldm | glrlm | glszm | ngtdm | shape |
|---|---|---|---|---|---|---|---|
| exponential | 0.16 |  | 0.06 | 0.08 | 0.07 |  |  |
| gradient | 0.29 | 0.28 | 0.20 | 0.18 | 0.24 | 0.28 |  |
| lbp | 0.08 |  | 0.06 | 0.08 | 0.07 |  |  |
| logarithm | 0.13 | 0.12 | 0.11 | 0.11 | 0.10 | 0.10 |  |
| original | 0.22 | 0.21 | 0.15 | 0.16 | 0.17 | 0.24 | 0.06 |
| square | 0.24 | 0.36 | 0.21 | 0.23 | 0.28 | 0.31 |  |
| squareroot | 0.18 | 0.16 | 0.14 | 0.14 | 0.13 | 0.18 |  |
| wavelet | 0.23 | 0.16 | 0.13 | 0.13 | 0.13 | 0.14 |  |

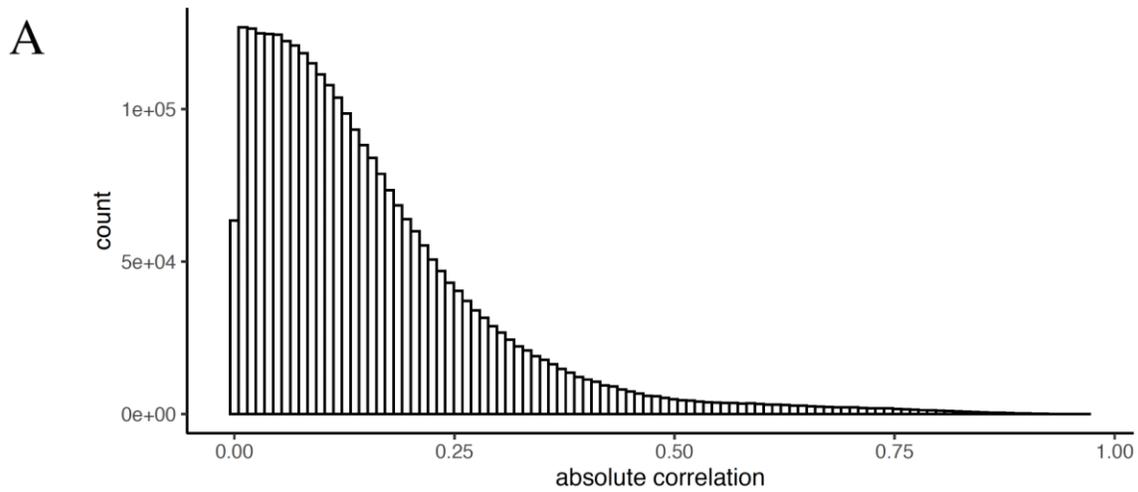

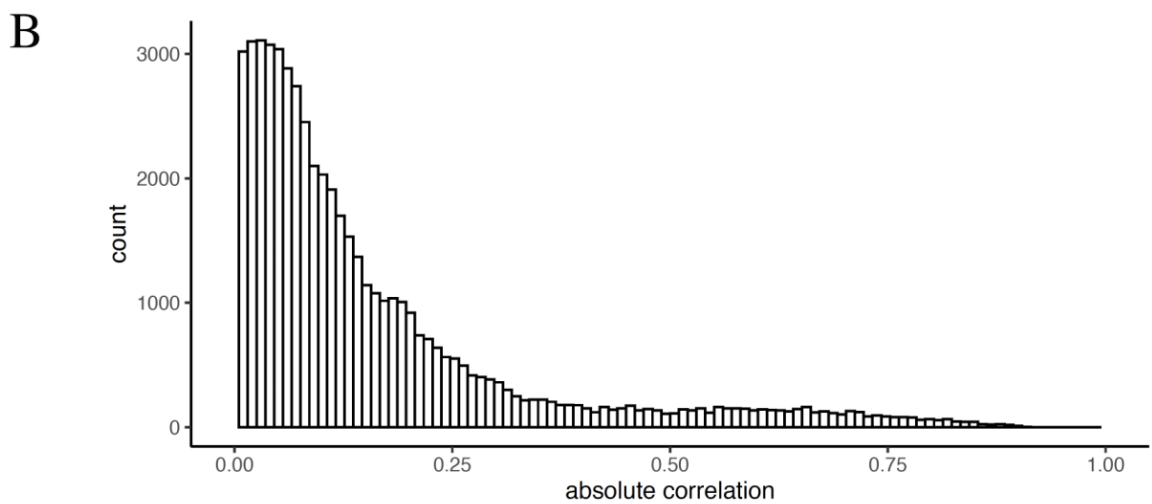

**Figure 3**. Histogram of absolute correlation coefficients

A. Correlation coefficients between PyRadiomics and ImgRes

B. Correlation coefficients between PyRadiomics and LungRes

**Prognosis Performance of the Proposed Prognosis Model**

As shown in Figure 4, the performances of three feature reduction methods (PCA, Boruta, and CPH) were compared to that of the proposed risk score-based prognosis model.

PCA method generated 44 components to represent the variance in the original 3,540 features of the combination of the PyRadiomics, ImgRes, and LungRes feature banks. Boruta feature reduction method selected 4 features in 1,000 iterations, with a cut off at 0.2 (p-value cut off for Boruta method). CPH method identified 325 features associated with overall survival in the training cohort. Particularly, 283 of them belong to the PyRadiomics feature bank, while 40 features were extracted using ImgRes. LungRes also contributed another 2 features. The proposed risk score-based model utilized the probabilities of the three individually trained Random Forests classifier based on the PyRadiomics, ImgRes, and LungRes feature sets. The AUC for each method was calculated on the test cohort.

The AUC for PCA, Boruta, and CPH methods were 0.65, 0.57, and 0.53, respectively. The proposed risk score-based method produced the highest AUC of 0.86. Comparing the feature reduction methods using specificity test, the performance of the proposed risk score-based method was significantly higher than PCA (p-value = 0.035), Boruta (p-value = 0.025), and Cox-Regression methods (p-value = 0.0031). The results suggest that an "ensemble" model, which is based on probabilities calculated by multiple individual small models, gave the best performance compared to other models. The ROC curves for three traditional feature reduction methods (PCA, Boruta, and CPH) and the proposed risk score-based model are shown in Figure 4.

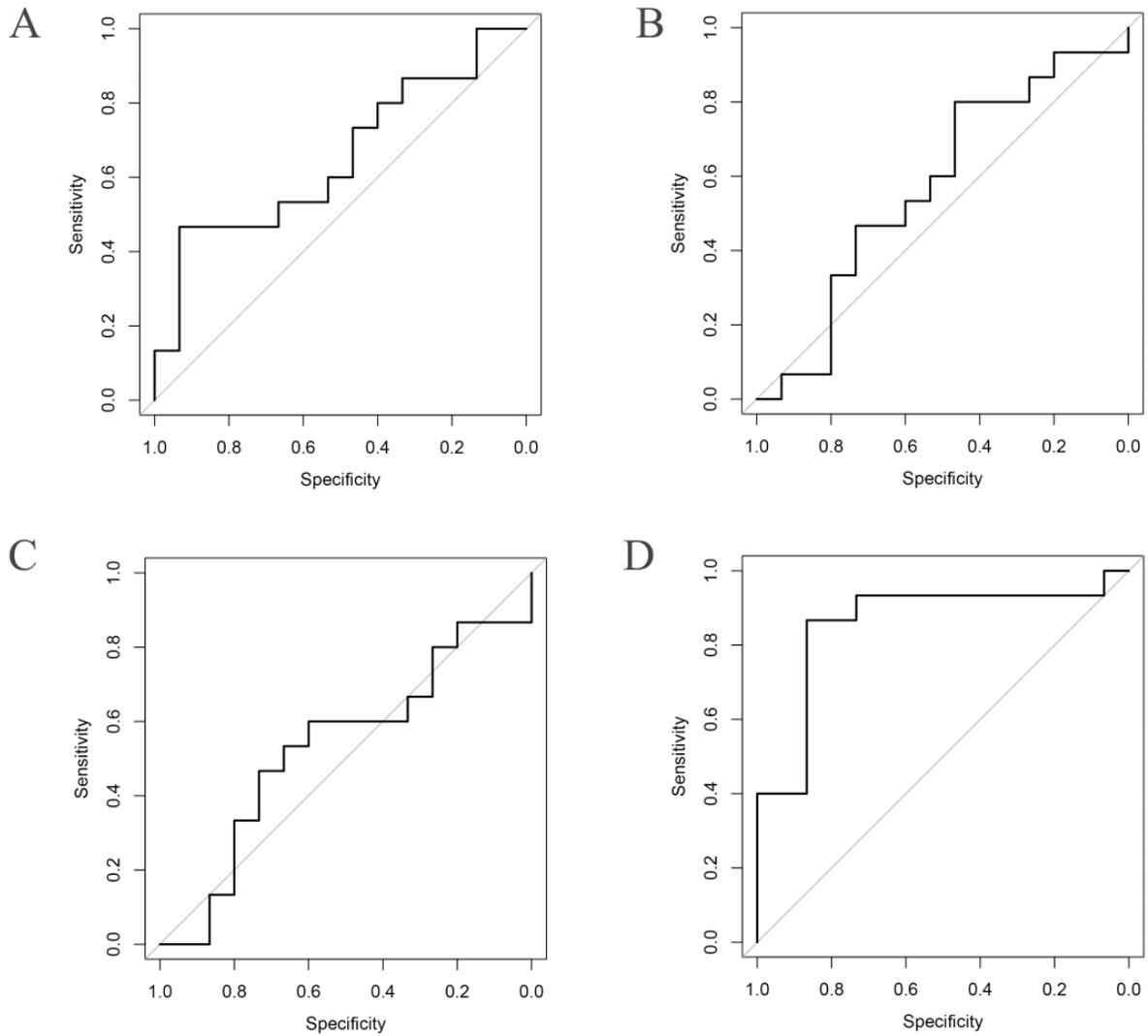

**Figure 4**. ROC curves of models using four feature reduction methods.

A. ROC curve for PCA based fusion method, AUC = 0.65.

B. ROC curve for Boruta based feature reduction method, AUC = 0.57.

C. ROC curve for CPH based feature reduction method, AUC = 0.53.

D. ROC curve for risk-score based feature fusion method, AUC = 0.86.

**Discussion**

In this study, we proposed a novel risk score-based feature reduction and fusion method for a prognosis model and compared it to three different feature reduction methods in PDAC CT image for predefined radiomics and deep transfer learning feature banks. We discovered that the proposed risk score-based method (ensemble model) had a better prognosis performance than those of traditional supervised and unsupervised methods increasing AUC by 32% (AUC of 0.86 vs 0.65). This result is consistent with previous studies that ensemble methods can outperform traditional machine learning models[52–54]. In a previous study, each of individual models based on PyRadiomics, LungRes and ImgRes transfer learning models was evaluated individually and the best performance was achieved by ImgRes model (AUC of 0.74)[34]. It is promising to observe that the proposed risk score-based ensemble model outperforms the best individual model by 16% (AUC of 0.86 vs 0.74). In addition, the proposed ensemble model outperforms the predefined radiomics (PyRadiomics) model by 51% (AUC of 0.86 vs 0.57)[34].

As transfer learning increasingly plays a vital role in medical image analysis, the curse of dimensionality is becoming more acute in radiomics-based prognosis models[1]. Supervised feature reduction methods such as univariate CPH and Boruta have difficulties in balancing false positive rate and statistical power. By testing 3,540 features using univariate CPH, the probability of having at least one false positive is higher than 99%. Hence, supervised feature reduction methods may lose their significances as feature banks continue to grow in size. In addition, unsupervised methods including PCA and Independent Component Analysis (ICA) were not able to boost the prognosis performance due to the inherent noise in images features. On the other hand, ensemble methods, which use multiple models to generate risk scores, may overcome these limitations of the traditional feature reduction methods[55,56]. Additionally, since risk scores were generated using the non-linear classifier Random Forest, they were non-linear mappings of the original feature space, providing better fits for patients' survival patterns. In our study, using resectable PDAC CT images, the proposed ensemble method had significantly higher AUC compared to other feature fusion and reduction methods including PCA (p-value = 0.035), Boruta (p-value = 0.025), and Cox-Regression (p-value = 0.0031).

It is worth to note that although there were high Pearson correlations coefficients between certain deep radiomics and PyRadiomics features, most deep radiomics features were independent from engineered PyRadiomics features. This result suggests that the relationship between deep radiomics and PyRadiomics was more complementary than replacement. Since most deep radiomics features do not have a linear relationship with engineered radiomics features, fusing these two feature banks would provide more information to the prognosis model. On the other hand, existing correlations between radiomics and deep radiomics features suggest that through backpropagation, pretrained CNNs were also able to capture similar patterns. Future study can further test the associations between conventional radiomics features and deep radiomics features. A thorough understanding of these associations will provide a steady base for developing more sophisticated and advanced feature fusion methods, which may further improve the performance.

Although the proposed ensemble method outperformed traditional approaches, it had limitations. Compared to supervised methods where certain biomarkers can be identified during the process, ensemble methods are hard to interpret since the stacked model is based on the results (probabilities) from other models. Although using intuitive algorithms such as logistic regression instead of Random Forests, the final prognosis probability (risk score) can be derived from original features using mathematical formulation, it would be a complicated task. In addition, the current models provide binary outcomes as final outputs, ignoring the time to the event. It would be better to include time duration to further improve the prognosis model.

**Conclusion**

Deep radiomics features are complementary to conventional radiomics features. Through the proposed risk score-based prognosis model by fusing deep radiomics and radiomics features, prognostication performance for resectable PDAC patients showed significant improvement compared to deep learning or predefined radiomics features alone.